# Brain2Text Decoding Model Reveals the Neural Mechanisms of Visual Semantic Processing

Feihan Feng[1-5] & Jingxin Nie[1-5*]

[1]Philosophy and Social Science Laboratory of Reading and Development in Children and Adolescents (South China Normal University), Ministry of Education Center for Studies of Psychological Application, South China Normal University; Guangzhou, 510631, China.

[2]Center for Studies of Psychological Application, South China Normal University; Guangzhou, 510631, China.

[3]Key Laboratory of Brain, Cognition and Education Sciences (South China Normal University), Ministry of Education.

[4]School of Psychology, South China Normal University; Guangzhou, 510631, China.

[5]Guangdong Key Laboratory of Mental Health and Cognitive Science, South China Normal University; Guangzhou, 510631, China.

*Corresponding author Email: niejingxin@gmail.com.

**Abstract**

Decoding sensory experiences from neural activity to reconstruct human-perceived visual stimuli and semantic content remains a challenge in neuroscience and artificial intelligence. Despite notable progress in current brain decoding models, a critical gap still persists in their systematic integration with established neuroscientific theories and the exploration of underlying neural mechanisms. Here, we present a novel framework that directly decodes fMRI signals into textual descriptions of viewed natural images. Our novel deep learning model, trained without visual information, achieves state-of-the-art semantic decoding performance, generating meaningful captions that capture the core semantic content of complex scenes. Neuroanatomical analysis reveals the critical role of higher-level visual cortices, including MT+ complex, ventral stream visual cortex, and inferior parietal cortex, in visual semantic processing. Furthermore, category-specific analysis demonstrates nuanced neural representations for semantic dimensions like animacy and motion. This work provides a more direct and interpretable framework to the brain's semantic decoding, offering a powerful new methodology for probing the neural basis of complex semantic processing, refining the understanding of the distributed semantic network, and potentially developing brain-inspired language models.

## 1. Introduction

Humans seamlessly navigate the world through semantic understanding, effortlessly transforming sensory experiences into meaningful concepts that underpin language, object recognition, and social interaction. Deciphering the underlying neural mechanisms of this ability remains a challenge in neuroscience. Functional neuroimaging studies have identified a left-lateralized semantic network comprising multiple brain regions involved in semantic processing(Binder, Desai, Graves, & Conant, 2009), broadly organized into two principal interacting neural systems: the representation system and the control system(Ralph, Jefferies, Patterson, & Rogers, 2017). Within this network, regions such as the middle temporal gyrus (MTG) and the anterior temporal lobe (ATL) are crucial for representing and integrating multimodal conceptual knowledge(Lau, Phillips, & Poeppel, 2008; Patterson, Nestor, & Rogers, 2007), while the inferior frontal gyrus (IFG) is associated with computing and manipulating activation in the representation system to suit the current context or goals(Davey et al., 2015). Alongside the substantial progress in mapping this network, a critical question emerges: how does the brain encode the rich and nuanced semantic information derived from complex, real-world experiences?

Early theories propose that semantic information may be represented in sparse and independent brain regions based on semantic categories(Frisby, Halai, Cox, Ralph, & Rogers, 2023), and certain regions exhibit specific responses to particular semantic categories, such as faces and buildings(Caramazza & Shelton, 1998; Epstein & Kanwisher, 1998; Kanwisher, McDermott, & Chun, 1997). However, it is unlikely that thousands of distinct categories are represented in distinct brain regions. Instead, conceptual representations may emerge from perceptual or motor representations distributed across different brain regions. For example, the concept of "tomato" is associated with color and shape, and "scissor" is linked to hand-related motion(Aziz-Zadeh, Wilson, Rizzolatti, & Iacoboni, 2006; Fernandino et al., 2016; Liuzzi, Aglinskas, & Fairhall, 2020; Tyler et al., 2013). Previous studies leveraging computational models have further delineated a distributed network encoding semantic knowledge of concepts and perceptual/motor features(Huth, De Heer, Griffiths, Theunissen, & Gallant, 2016; Huth, Nishimoto, Vu, & Gallant, 2012; Mitchell et al., 2008). Classical approaches often deliberately rely on linear encoding models, because they provide an elegantly interpretable framework for characterizing neural representation structure. However, linear models may not fully capture the complex, nonlinear neural dynamics in semantic processing. Moreover, many studies utilize linguistic stimuli, neglecting both the neural mechanisms underlying multimodal semantic encoding and the ecological validity in natural scenes.

Vision, as our dominant sensory modality, provides an unparalleled source of semantic information. Natural images, in particular, provide more diverse and complex stimuli that elicit ecologically valid neural responses compared to linguistic symbols, offering a powerful avenue for investigating semantic understanding(Hasson, Malach, & Heeger, 2010). Early efforts to decode perceptual experiences from brain activity primarily focused on reconstructing low-level visual features such as edges, colors, and textures, producing outputs devoid of interpretable semantic information(Shen, Dwivedi, Majima, Horikawa, & Kamitani, 2019; Shen, Horikawa, Majima, & Kamitani, 2019). Recent studies combining large-scale datasets recording fMRI signals as participants viewing natural images, such as the Natural Scenes Dataset (NSD)(Allen et al., 2022), with multimodal language models such as CLIP(Radford et al., 2021) have improved the visual realism and semantic fidelity of reconstructed images(Chen, Qing, Xiang, Yue, & Zhou, 2023; Ferrante, Ozcelik, Boccato, VanRullen, & Toschi, 2023; S. Lin, Sprague, & Singh, 2022; Mai & Zhang, 2023; Ren et al., 2024; Takagi & Nishimoto, 2023). However, much of this progress has been driven by architectural advances rather than deeper exploitation of neural data, leaving the critical question of how humans encode visual input and transform it into meaningful semantic interpretaion elusive.

Meanwhile, a growing body of studies has demonstrated the feasibility of mapping neural signals onto purely semantic representations and decoding directly into text, which provides a more straightforward and interpretable window into the brain's visual semantic encoding mechanisms compared to visual reconstruction approaches. Luo et al. proposed a framework that generates natural-language descriptions for images predicted to maximally acitivate individual voxels of interest, revealing the voxel-level semantic perferences in higher-level visual regions(Luo, Henderson, Tarr, & Wehbe, 2023). Doerig et al. reconstructed text embeddings predicted from neural signals directly into text using nearest neighbor search method, proving that the embeddings of scene captions produced by Large Language Models (LLMs) can characterize brain activity evoked by natural scenes(Doerig et al., 2025). Horikawa's mind captioning generates text embeddings by averaging LLM-tokens of scene captions and achieve text reconstruction through iterative optimization(Horikawa, 2025). Nevertheless, these approaches are explaining neural mechanisms at the level of latent embeddings rather than at the level of concrete textual meaning, and how distributed cortical networks quantitatively contribute to visual semantic decoding remains obscured within complex, nonlinear architectures.

Here, we propose a novel framework that directly decodes semantic content of viewed natural images from fMRI signals (Fig. 1). Our model consists of two parts, an nonlinear encoder that transforms fMRI signals into sentence-level embeddings, and a decoder called Vec2Text(Morris, Kuleshov, Shmatikov, & Rush, 2023) that generates textual descriptions from these neural-derived representations. By excluding visual information from both the input data and target vectors and translating fMRI signals directly into textual descriptions, we aim to tap into the neural processes underpinning abstract semantic representations.

Additionally, encoding of integrated semantic concepts in sentence-level text embeddings more closely mirrors the brain's encoding strategy than does the word-by-word generation typically employed in language models(Lu, Jin, Ding, & Tian, 2023). Moreover, through SHAP analysis(Lundberg, 2017) we quantify the contributions of specific regions of interest (ROIs), identify brain regions critical for visual semantic processing, and characterize brain activation patterns associated with different semantic contents. In conclusion, our contributions are summarized as followed:

- We propose a model that directly decodes semantic content of natural images from brain signals without leveraging visual information, mimicking the brain's semantic processing from sensory input to abstract concepts.
- We introduce an interpretation pipeline based on SHAP analysis for a nonlinear decoding architecture, quantifying the marginal contribution of visual regions to semantic decoding and identifying the critical role of higher-level visual cortex in representing specific semantic.
- Our work provides a more interpretable window into visual semantic processing by grounding model interpretation in concrete reconstructed text, bridging the gap between decoding performance and neuroscientific insight.

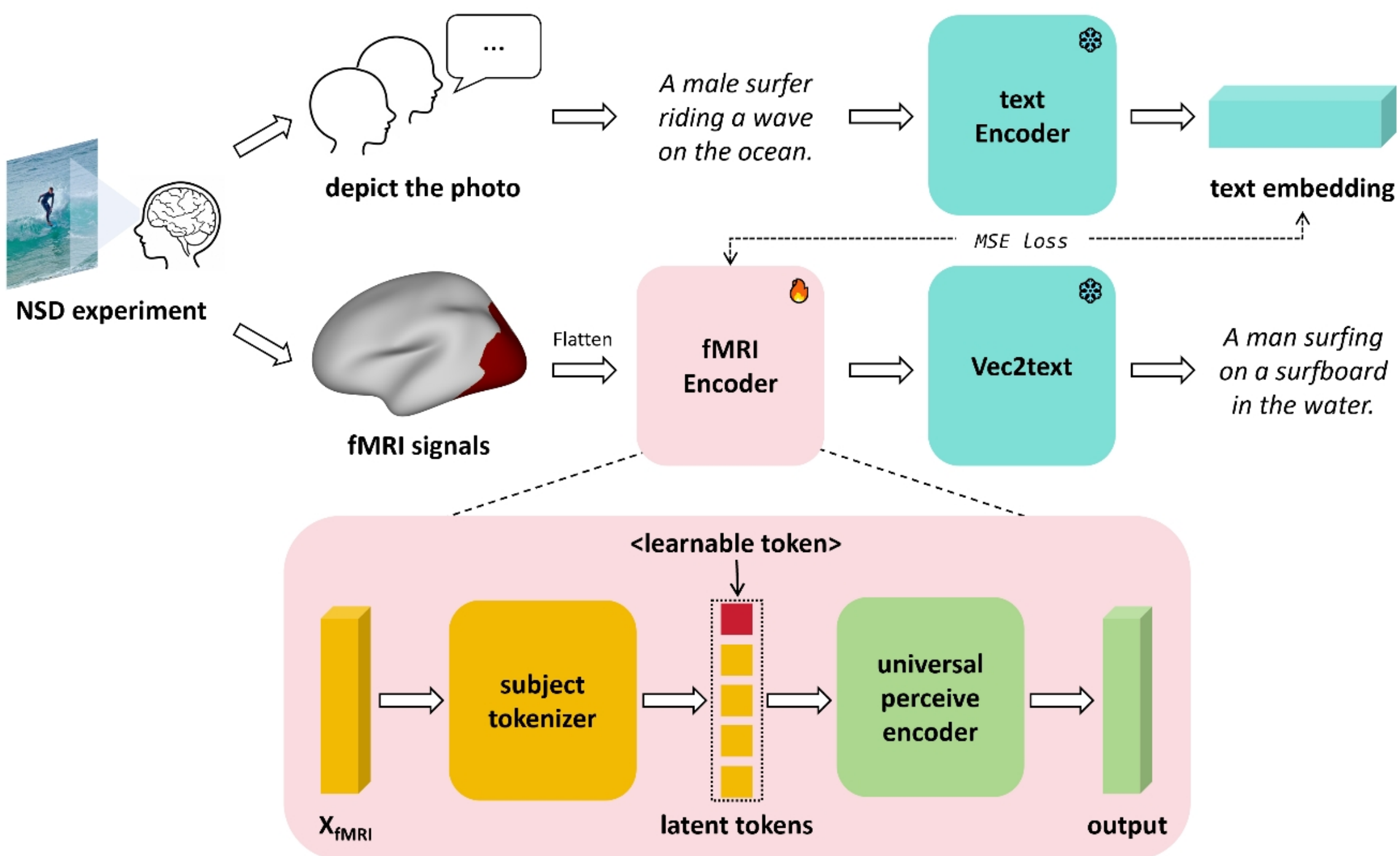

**Fig. 1.** The overview of our research.

## 2. Methods

### 2.1. Dataset

We utilized the Natural Scenes Dataset (NSD), a large-scale, high-quality fMRI dataset acquired from a 7-Tesla fMRI scanner across 30-40 sessions, during which each subject viewed 10,000 images from the COCO dataset(T.-Y. Lin et al., 2014), with each image repeated three times. Each image is annotated with five captions and multiple labels spanning 80 categories. Our analysis included data from four subjects (subj01, subj02, subj05, subj07) who completed all scanning sessions. Each subject viewed 9,000 unique images (27,000 trials) and 1,000 shared images (3,000 trials). All trials corresponding to unique images were included as training data. Test data were constructed by averaging the three trials of each shared image, resulting in a test set of 1,000 averaged responses of each image, which was held consistent for all four subjects. As a final preprocessing step for model input, the fMRI voxel data for each sample were flattened into a one-dimensional vector.

## 2.2. Model architecture

Our model consists of two parts: an encoder transforming fMRI signals into text embeddings, and a decoder reconstructing brain captions from the resulting latent vectors. The encoder, adapted from an established method(Xia, de Charette, Oztireli, & Xue, 2024), has demonstrated the ability to predict image embeddings from fMRI signals and perform downstream tasks such as image captioning, grounding, and visual decoding through the utilization of multimodal Large Language Models. Also, the cross-subject joint training architecture enhances both performance and generalization capabilities.

The brain encoder comprises two main components: (1) a subject-specific tokenizer that projects fMRI signals into a sequence of brain tokens. Given an input fMRI vector $x \in \mathbb{R}^{V_s}$ from subject $s$, the input vector is passed through two linear projection layers, one of which is followed by Layernorm, GELU activation function and Dropout (dropout rate = 0.5). This produces a latent sequence of shape $N \times L$, where $N$ denotes the dimension of text embedding ($N$ = 1536) and $L$ denotes the number of latent tokens. Next, the latent sequence incorporates a learnable subject-specific token of size $N$ that independent of the inputs. (2) a universal lightweight transformer architecture that uses cross-attention mechanism to project the brain tokens into a latent bottleneck, extracting subject-invariant semantic knowledge from the tokenized fMRI representation. Finally, a 1×1 convolution layer with $L$+1 input channels and 1 output channel is applied to aggregate information across latent channels and outputs a 1536-dimensional text vector for each sample, which is subsequently passed to the text decoder.

The decoder is a pretrained model that iteratively corrects and re-embeds text based on a fixed point in latent space(Morris et al., 2023). This model was trained to invert text embeddings from embedding models, and it indicated a sort of equivalence between raw text and its embeddings, which enabled direct decoding of raw text from the text embeddings. Meanwhile, due to the semantic properties of the text embeddings, similar texts exhibit consistent vector directions, allowing the inversion model to preserve the original semantics, even when averaging these embeddings. We use the text-embeddings-ada-002 model from OpenAI to encode image captions into vectors, which serve as ground truth embeddings without visual information. During decoding, we adopted the default setting described in the original document and performed 20 iterative correction steps to reconstruct textual descriptions corresponding to the images viewed by participants from brain-predicted embeddings.

Our model includes two key hyperparameters: the number of latent channels of the subject-specific tokenizer ($L$) and the depth of the universal transformer ($D$). In most hyperparameter configurations, the model's performance approached the training limit, indicating exhibition of overfitting. To mitigate this issue while optimizing the balance between performance, training efficiency, and model complexity, we set the number of latent channels to 4 and the transformer depth to 4 (Appendix E).

## 2.3. Training strategy

Our model was trained on a single NVIDIA RTX 4090 GPU for 200 epochs with a batch size of 256 per subject, totaling 1024 samples across four subjects. Prior to formal training, preliminary experiments were conducted using the validation set containing 300 images from subj01 to determine the optimal hyperparameters, and the held-out set of 1,000 shared images was reserved exclusively for final evaluation. We employed AdamW(Loshchilov, 2017) as the optimizer with β1 = 0.9, β2 = 0.95, and weight decay factor of 0.01. A one-cycle learning rate schedule(Smith & Topin, 2019) was used, with an initial and maximum learning rate of $1e^{-3}$. Model training was halted, and the best model checkpoint was saved once overfitting occurred. Let $\hat{y}_i \in \mathbb{R}^D$ denote the predicted text embedding and $y_i \in \mathbb{R}^D$ the target text embedding for the $i$-th sample (D = 1536). For a minibatch of size N, Mean Squared Error (MSE) weighted by a cosine similarity-based coefficient was served as the loss function:

$$loss = \frac{1}{N}\sum_{i=1}^{N}\left(2 - \frac{1}{2}\cos_sim(\hat{y}_i, y_i)\right)\|\hat{y}_i - y_i\|^2 \quad (1)$$

To enhance model generalization, one of the five captions for each ground truth image was randomly selected as the reference during training.

## 2.4. ROI selection

Previous methods based on the NSD dataset employed voxels from the "nsdgeneral", which indicates the occipital regions that are generally responsive in the NSD experiment, as model inputs. Here we introduced the *stream* mask based on Wang's anatomical atlas of visual topography(Wang, Mruczek, Arcaro, & Kastner, 2015) in the dataset to identify hierarchical visual ROIs: the lower-level ROIs, including the early visual cortex and the intermediate

ROIs, and the higher-level ROIs. Neuroscientific research has revealed the functional hierarchy within different levels of visual cortex(Grill-Spector & Malach, 2004). Additionally, visual decoding models also found that incorporating higher-level visual regions produces superior results compared to relying solely on the early visual cortex(S. Lin et al., 2022; Takagi & Nishimoto, 2023). Therefore, we excluded low-level visual features and used the higher-level ROIs as model inputs to enhance semantic reconstruction performance. To further analyze the functional roles of different regions in semantic processing, we parcellated the higher-level ROIs using the HCP_MMP atlas(Glasser et al., 2016), identifying approximately 40 regions that were grouped into nine ROIs after excluding regions with insufficient voxels (fewer than 10 voxels in 1.8mm native volume space, Appendix F). These included: the secondary visual cortex region, ventral stream visual region, dorsal stream visual region, MT+ complex region, medial temporal region, lateral temporal region, TPO (temporo-parieto-occipital) junction region, superior parietal region and inferior parietal region.

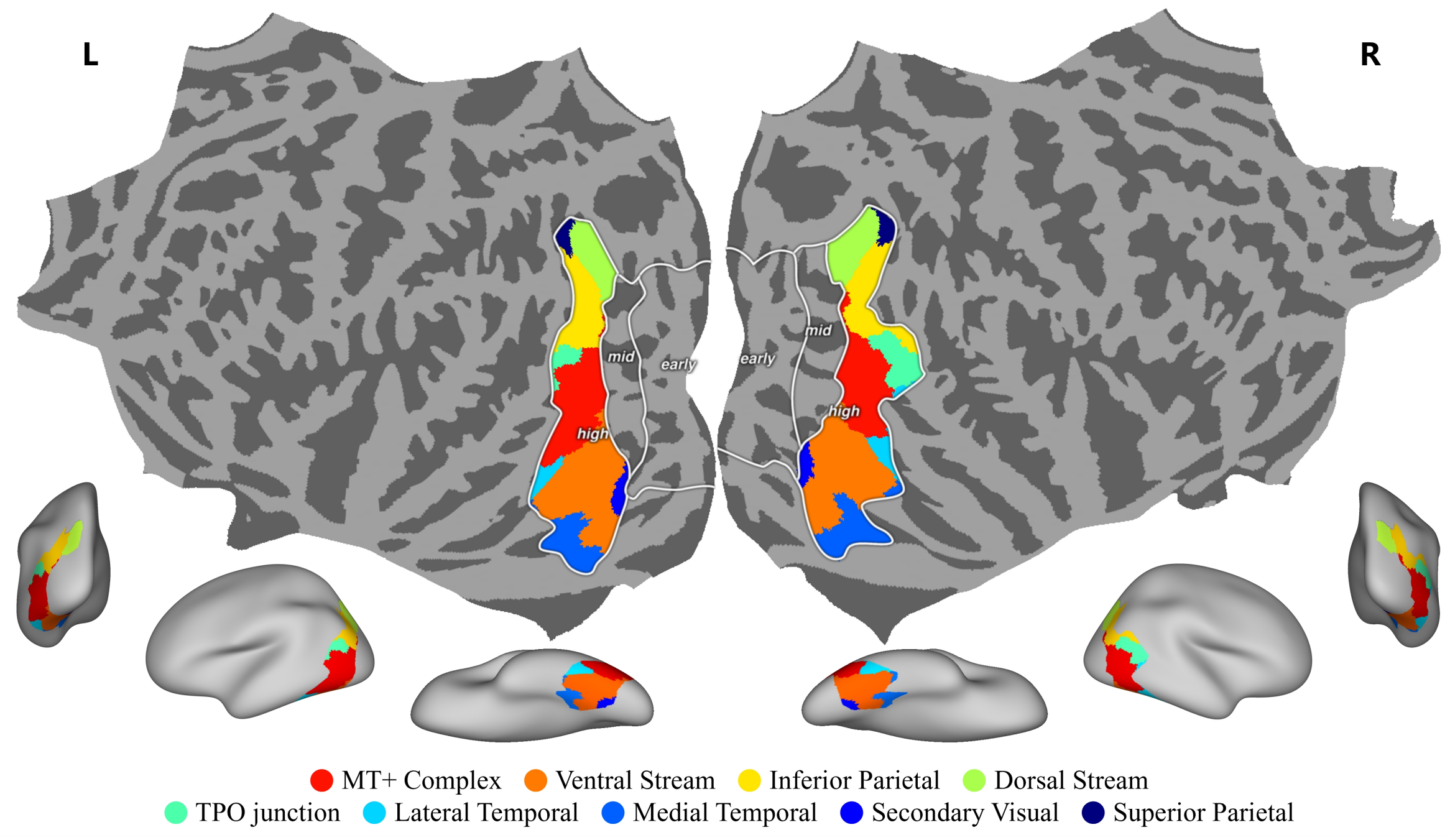

**Fig. 2.** This flattened cortex shows the parcellation of input voxels. The white contour delineates the "nsdgeneral" mask, representing visual cortex voxels conventionally used as model input in previous methods. The colored areas correspond to higher-level visual regions identified through the HCP_MMP atlas. ('early', early visual cortex ROI; 'mid', the intermediate ROIs; 'high', the higher-level ROIs)

## 2.5. Evaluation metric

To evaluate our model's performance, we selected five established Natural Language Processing (NLP) metrics: BLUE-k(Papineni, Roukos, Ward, & Zhu, 2002), METEOR(Banerjee & Lavie, 2005), ROUGE(C.-Y. Lin, 2004), CIDEr(Vedantam, Lawrence Zitnick, & Parikh, 2015) and SPICE(Anderson, Fernando, Johnson, & Gould, 2016), to quantify the semantic and syntactic similarity between reconstructed text and ground truth captions. Additionally, we incorporated two CLIP-based metrics, CLIP-S and RefCLIP-S(Hessel, Holtzman, Forbes, Bras, & Choi, 2021), to evaluate the alignment between reconstructed text and both ground truth images and captions. For ROI analysis, input voxels were aggregated into functional relevant brain regions to reduce computational complexity while preserving neurobiological interpretability. We applied SHAP values(Lundberg, 2017) to quantify the average marginal contribution of each feature across all possible subsets, providing an unbiased estimate of feature importance. Here, CLIP-S served as the evaluation metric for SHAP analysis, ensuring that the assessed contributions directly reflected alignment with visual-semantic representations. To provide a reference upper bound for our framework, we evaluated test set images using their ground truth captions. Specifically, text reconstructed from the average of their embeddings instead of a specific caption was evaluated using the aforementioned metrics, representing the theoretical maximal performance achievable on the current framework and evaluation pipeline.

## 2.6. SHAP analysis

To quantify regional contributions to semantic decoding, we perform a ROI-level attribution analysis through a SHAP-based attribution analysis. We treated each ROI defined in the previous section as a feature group in the SHAP framework. Specifically, let $M = 9$ denotes the number of ROI features. For any given coalition (subset) of $S \subseteq \{1, \cdots, M\}$, we constructed a masked fMRI input by retaining voxel values within the ROIs included in $S$ and replacing voxels outside $S$ with the background data, which computed from the average activation pattern across the training set. The masked input was then fed into the frozen brain encoder and the pretrained text decoder to generate textual captions. We defined the scalar model output for each coalition as the CLIP-S between reconstructed caption and the corresponding image. The SHAP value for ROI $i$ was then estimated using the Shapley formulation:

$$\varphi_i = \sum_{S \subseteq \{1, \ldots, n\} \setminus \{i\}} \frac{|S|!\,(n - |S| - 1)!}{n!} [f(S \cup \{i\}) - f(S)] \tag{2}$$

where $f(S)$ denotes the CLIP-S obtained from the masked input containing only ROIs in subset $S$. In practice, the final ROI contribution was summarized by averaging SHAP values across test set. And a positive SHAP value indicates that retaining the voxel values from that ROI (relative to replace them with background data) increases the semantic similarity between reconstructed captions and the corresponding images.

## 2.7. Category analysis

The images utilized in the NSD experiment, sourced from the COCO dataset, are annotated with multiple semantic labels spanning 80 categories. To systematically evaluate the semantic content of the text reconstructed by our model, we employed the Large Language Model (Llama3-8B) for automated classification, as LLMs have demonstrated human-comparable or superior performance on various cognitive tasks, including text classification and multimodal processing(Binz & Schulz, 2023). In our study, we set the prompt as: "Is [category] included in the content of this sentence? Answer with only yes or no: ", followed by the reconstructed semantic description derived from fMRI signals. This process enabled us to determine whether each reconstructed text could be assigned to any of 80 semantic categories. Additionally, images associated with labels such as person or animals were grouped into the "living" category, while the remaining images were classified as "non-living". The "non-living" group was further subdivided based on the presence of motion-related objects or scenes (e.g., "surfboard", "bicycle").

To explore the contribution differences of ROIs between semantic categories, we conducted a covariate-adjusted permutation test(Winkler, Ridgway, Webster, Smith, & Nichols, 2014), controlling overall decoding performance (CLIP-S), caption length as the number of words after lowercasing and removing punctuation, and lexical diversity as type-token ratio (TTR). Specifically, we first fit a full model that contains both the effect of interest (category label) and the nuisance variables and computed the statistic of interest $T_0$. Next, we fit a reduced model containing only the nuisance covariates, obtained its fitted values (the estimated nuisance effects) and residuals, and then permuted the residuals 10,000 times. For each permutation, the permuted residuals were added back to the fitted values from the reduced model to generate a permuted response vector, and the full model was refit to this permuted data, computing the statistic of interest $T_j$ similarly and constructing a null distribution of T values. Two-sided p values were computed by counting how many times the absolute $T_j$ was found to be equal to or larger than absolute $T_0$, and divided the count by the number of permutations. At last, Benjamini-Hochberg FDR correction was applied to the resulting ROI-level p values.

## 3. Results

### 3.1. Evaluation of the model's performance

Fig. 3 presents examples of image captions reconstructed from fMRI signals by our model alongside human-provided reference captions. Although exhibiting occasional inaccuracies in fine-grained details and minor syntactic errors, the reconstructed captions show robust semantic fidelity, effectively capturing core content of corresponding images. This suggests that our model is capable of decoding visual semantics from brain activity without visual input. In order to evaluate the representativeness of the displayed samples in the test set, the distribution of CLIP-S scores for all reconstructed captions across the four participants is shown in Appendix A. The examples presented in Fig. 3 are highlighted within this distribution.

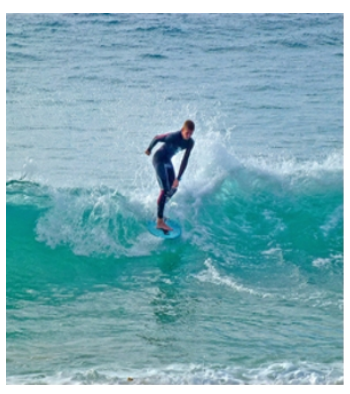

A man surfing on a surfboard in the water.
A man surfing on a surfboard in the water.
A man on a surfboard is riding in the water.
A man riding a surfboard in the ocean.
A person on the surf is riding a wave at the ocean.

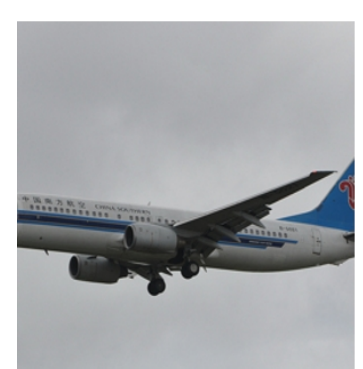

A blue plane flying in the sky on a large jet.
A large airplane flying in the sky.
A plane with a large jet aircraft is sitting on the runway in blue skies.
A large blue airplane flying in the sky through the air.
A large Chinese passenger plane is flying through the cloudy sky.

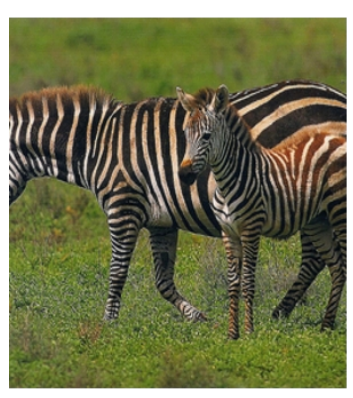

Two zebras standing in a field with baby zebras walking across the grass.
Two zebras are standing in a field of grass walking next to each other.
Two zebras standing in a herd walking along a patch of grass.
Two giraffes. Two zebras are standing next to each other in the grass.
A couple of zebras standing in a field on the grass.

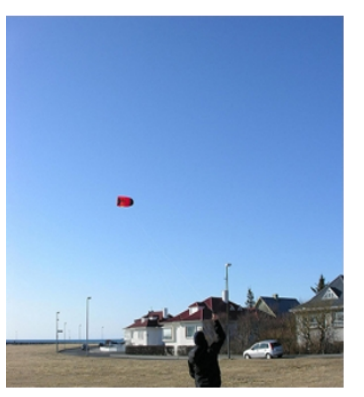

A person flying a kite in the sky is standing on a man's pole.
A man flying kites in the sky over a couple of people with a kite on his pole.
A person flying a kite in the sky on a green field.
A man flying a kite on a plane in the grass with his palms against the ocean.
A man is flying a kite in the sand on a field beside the bare ground.

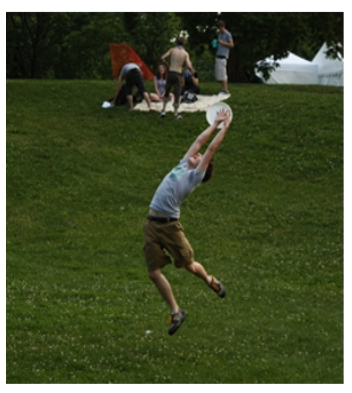

A man in the grass throwing a yellow frisbee on a field.
A baseball player is swinging a ball at a pitcher.
A man in a yellow shirt catching a Frisbee in the air.
A man throwing a frisbee in the air.
A man is jumping to catch a frisbee on a blanket with two people out in the air.

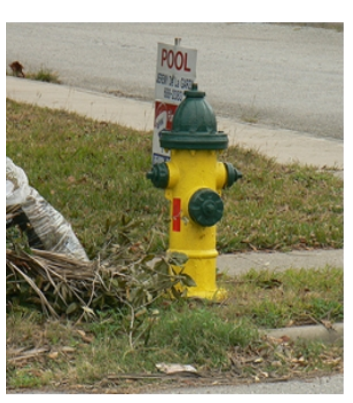

A red fire hydrant with a yellow stop sign sitting on a grass in the middle.
A red hydrant with a yellow stop sign on a fence parked in the grass.
A fire hydrant with a yellow and red spray sits on the street next to a dirt road.
A red and yellow fire hydrant on a sidewalk next to a brick stop sign.
A fire hydrant on the corner of a street with a yellow and green sign.

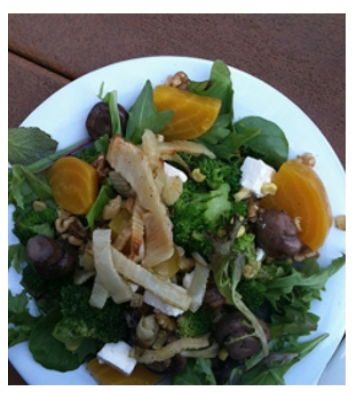

A plate of vegetables with a scallions and meat on it topped with lettuce and green cheese.
A bowl of sliced vegetables with a mixture of broccoli, red and orange meat on the table.
A plate of food with a bowl of vegetables, broccoli and red meat on it.
A plate of food with a piece of broccoli, greens and white rice on it.
A salad with greens, mushrooms and type of cheese smothered in a white plate.

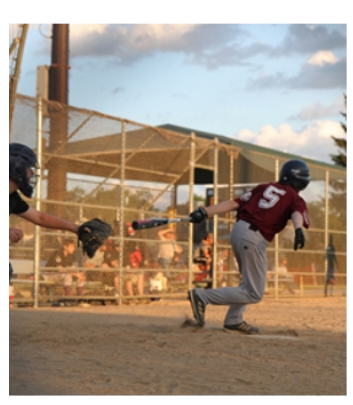

A man and two girls playing baseball in a field with a frisbee thrown on the grass.
A man hitting a baseball on a court with his bats in the air.
A man swinging a baseball at a pitcher in the grass with his fists on the disc.
A baseball player on a ring with a pitcher batting against the ball.
A baseball player with a bat in his hand and a batter standing on the plate in a game.

**Fig. 3.** Examples of reconstructed captions and their corresponding images. Colors denote the reconstruction results from different subjects: red for subj01, blue for subj02, green for subj05, and orange for subj07. The black text represents the averaged result reconstructed from the human-given caption embedding.

To optimize the model architecture and find out the optimal model parameters, we first conducted hyperparameter tuning using fMRI data from a representative participant (subj01) and evaluated decoding performance using CLIP-S. This analysis demonstrated that semantic information could be effectively decoded with a lightweight minimalist model architecture, while excessive parameterization tended to result in overfitting and led to performance degradation. Based on these findings, we established an optimal hyperparameter configuration with latent channels of 4 and the transformer depth of 4 (Appendix E).

We then benchmarked our model against several other brain decoding models, which all utilized visual cortex voxels data from subj01(Ferrante et al., 2023; Han et al., 2024; Takagi & Nishimoto, 2023; Xia et al., 2024). Notably, in contrast to prior state-of-the-art models that relied on image embeddings as intermediate representations, our model achieved comparable performance by direct decoding without leveraging visual input (Table 1). Furthermore, the model trained on higher-level ROIs significantly outperformed models using all visual cortex voxels or those focused on lower-level visual ROIs. SHAP analysis further implicated that higher-level visual cortex holds a dominant role in encoding semantic content of visual stimuli (Appendix B). Consequently, utilizing voxels from higher-level visual cortex as model inputs for semantic decoding proves more advantageous and yields superior performance. In addition, consistent with the notion that data quality is paramount in neuroimaging, evaluations across participants revealed that enhanced decoding performance is associated with high-quality fMRI data(Appendix C).

**Table 1** Evaluation results

| Method | BLEU1 | BLUE2 | BLEU3 | BLEU4 | METEOR | ROUGE | CIDEr | SPICE | CLIP-S | RefCLIP-S |
|---|---|---|---|---|---|---|---|---|---|---|
| Upper Bound | 87.07 | 69.31 | 52.28 | 38.80 | 36.98 | 61.97 | 136.78 | 32.10 | 80.27 | 85.74 |
| SDRecon | 36.21 | 17.11 | 7.72 | 3.43 | 10.03 | 25.13 | 13.83 | 5.02 | 61.07 | 66.36 |
| OneLLM | 47.04 | 26.97 | 15.49 | 9.51 | 13.55 | 35.05 | 22.99 | 6.26 | 54.80 | 61.28 |
| BrainCap | 55.96 | 36.21 | 22.70 | 14.51 | 16.68 | 40.69 | 41.30 | 9.06 | 64.31 | 69.90 |
| UMBRAE | 57.84 | 38.43 | 25.41 | 17.17 | 18.70 | 42.14 | 53.87 | 12.27 | 66.10 | 72.33 |
| Ours (Higher-level ROIs) | 56.79 | 36.47 | 21.85 | 13.07 | 21.56 | 43.20 | 47.84 | 14.01 | 69.14 | 74.07 |
| Ours (All ROIs) | 54.94 | 35.28 | 21.06 | 12.41 | 20.61 | 41.83 | 42.02 | 13.03 | 67.90 | 72.56 |
| RidgeCV | 44.29 | 24.33 | 11.17 | 5.25 | 14.48 | 34.74 | 10.72 | 5.43 | 57.28 | 60.35 |

Evaluation results of our model and other method. 'Higher-level ROIs' refers to the model trained with brain regions of higher-level visual cortex. 'ALL ROIs' refers to the model trained with same input data as methods above. 'RidgeCV' refers to a linear baseline trained with voxels in higher-level ROIs. 'Upper Bound' refers to the theoretical upper bound of model's performance. The color red represents the best performance, orange indicates the second-best, and yellow denotes the third-best. Our model achieved not only the best performance on high-level metrics, such as CLIP-S, but also comparable performance on low-level metrics, such as BLEU-k scores.

### 3.2. Interpreting the Neural Basis of Semantic Processing

To dissect the neural mechanisms of semantic processing, we leveraged SHAP values to quantify the contribution of different brain regions to the model's performance. Fig. 4 presents the contribution of different brain regions to the CLIP-S in our model. The MT+ complex, ventral stream visual and inferior parietal regions make substantial contributions to the model's performance, suggesting that these regions represent majority of semantic information in visual semantic processing. In contrast, the dorsal stream visual and TPO junction regions exhibit relatively lower contributions, suggesting the potential involvement in encoding domain-specific semantic content or higher-order semantic integration. Other regions may play a role unrelated to representing semantic information in semantic processing or may not be involved in semantic processing, instead reflecting the model's attribution tendencies across input features. As shown in appendix B, although lower-level ROIs contribute positively to model's performance, excluding these regions from model inputs still enhances the model's performance.

To assess the generality of ROI contributions in semantic processing, we conducted a supplementary validation on an independent dataset BOLD5000(Chang et al., 2019) (details in Appendix D). The main pattern of ROI contributions was broadly consistent with the NSD results, supporting the cross-dataset robustness of observed regional hierarchy.

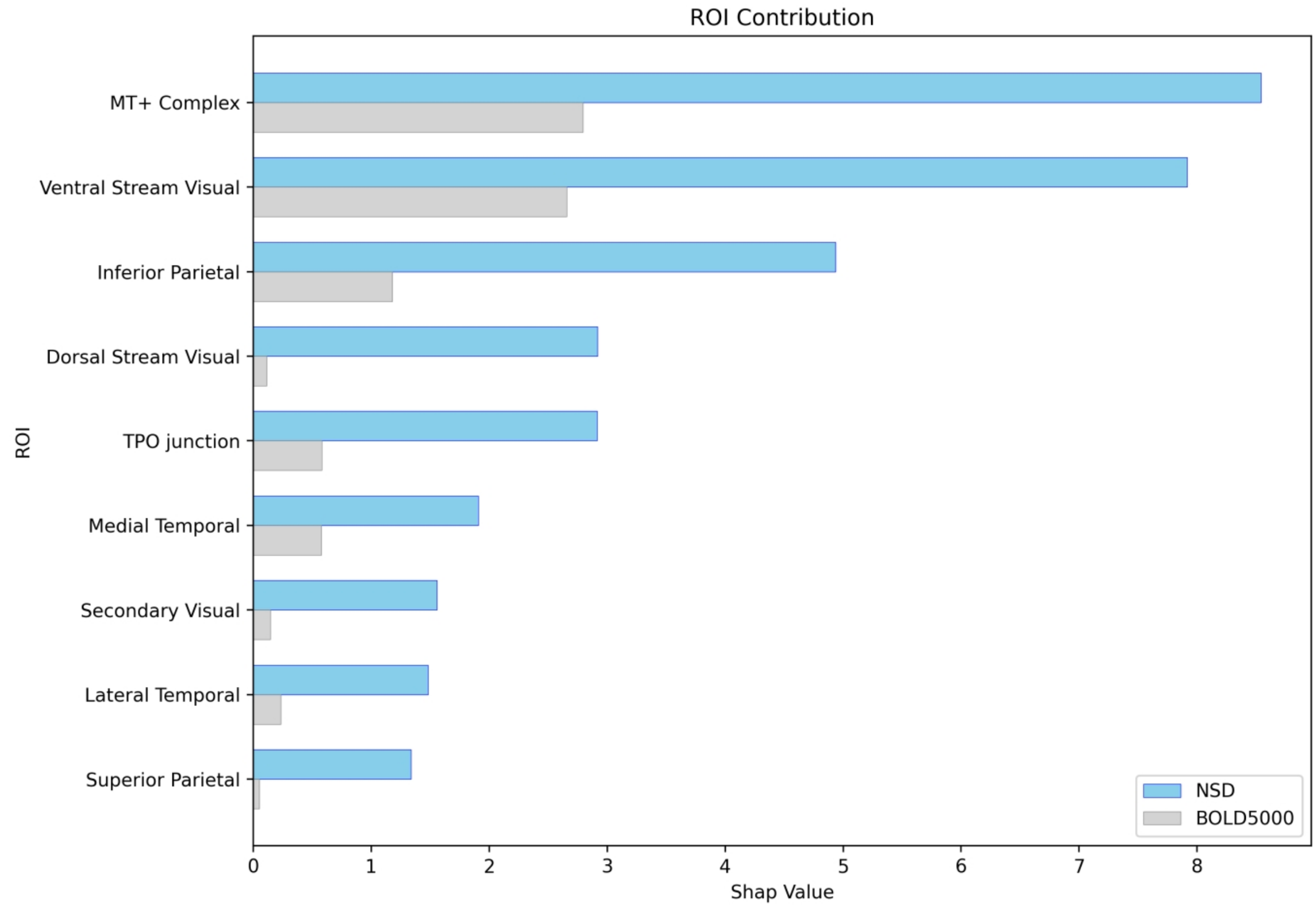

**Fig. 4.** The contribution of distinct brain regions in higher-level visual cortex to visual semantic decoding performance using SHAP values.

### 3.3. Category-specific semantic decoding

To investigate the neural encoding of diverse semantic categories, we evaluated the model performance across 80 distinct semantic categories using LLM. Fig. 5 presents the classification results evaluated with F1 score, upper bound was conducted to indicate the theoretical upper bound of model's performance. Our model effectively decodes semantic information containing the category "person", which aligns with the functional specialization of brain regions (e.g., FFA, EBA) in mediating selective recognition of human features(Downing, Jiang, Shuman, & Kanwisher, 2001; Kanwisher et al., 1997). It also demonstrates reasonable performance in reconstructing semantic information for motion-related concepts (e.g., "surfboard"), animals (e.g., "giraffe", "zebra"), and scenes involving "airplane" and "train". However, as expected, decoding accuracy varied across categories, with certain object categories (e.g., "backpack", "toaster") proving more challenging to accurately decode. This variability may reflects differences in neural representation strength, perceptual saliency, or the complexity of semantic features associated with different categories.

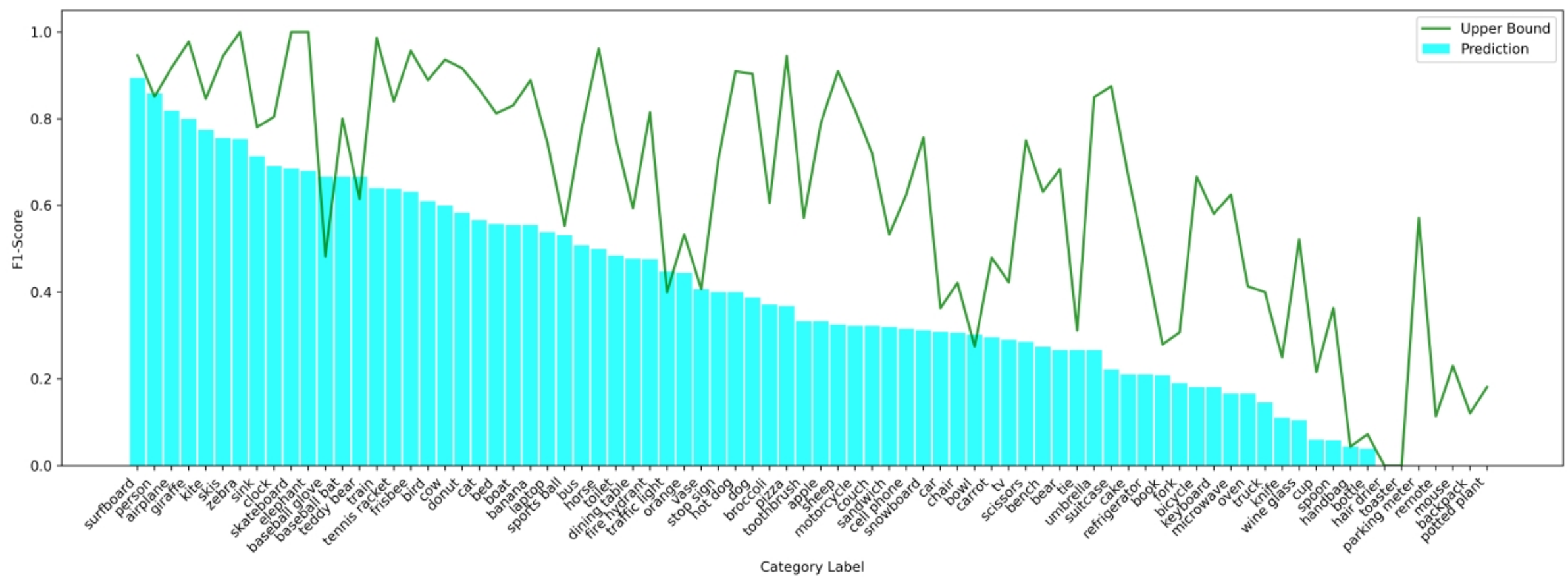

**Fig. 5.** Performance in category-specific classification based on reconstructed text, reflecting the model's ability to decode distinct semantic content that categorized according to 80 labels of ground truth images in the COCO dataset.

To analyze the relationship between semantic categories and different brain regions, we classified reconstructed text into "living" and "non-living" categories based on ground truth semantic labels, with non-living items subdivided into two groups defined by the presence of motion-related objects and scenes (e.g., "surfboard"). Permutation tests were applied to reveal differences in the SHAP values across five ROIs that contributed markedly to decoding. After adjusting overall decoding performance, caption length and linguistic variability, the "living" category exhibited significantly lower SHAP values in the MT+ complex, ventral stream visual and inferior parietal regions, and higher SHAP values in dorsal ventral stream and TPO junction (Fig. 6a), even though there are significantly lower CLIP-S for the "living" category (M = 68.5, SD = 12.4) compared to the "non-living" category (M = 70.3, SD = 11.3) in overall decoding performance ($\Delta$ = -1.81, $p$ = 0.04) (Fig. 6a). The motion-related items are associated with significantly higher SHAP values in the MT+ complex, ventral stream visual and dorsal stream visual regions, although motion-related items showed significantly reduced lexical diversity relative to "non-motion" items ($p$ < 0.001), however there are no significant differences in CLIP-S compared to "non-motion" items (Fig. 6b).

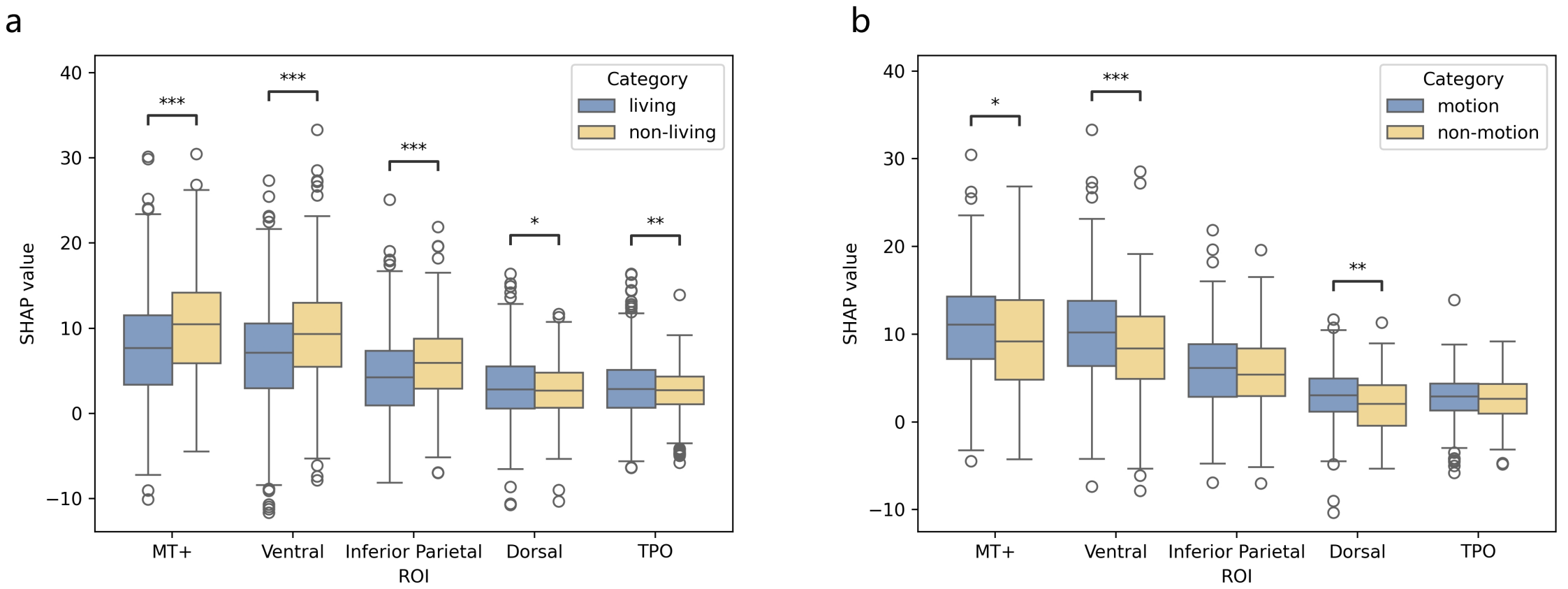

**Fig. 6.** SHAP values reveal distinct contributions of brain regions to visual semantic processing (higher SHAP values indicating greater involvement in the corresponding semantic categories). (a) Difference in contribution across five ROIs between images with living (human and animals) and those without living content. (b) Difference in contribution across five ROIs within non-living images, revealing the difference of images with and without motion-related objects and scenes. Statistical significance was assessed using

permutation tests (n=10,000). (*, $p < 0.05$; **, $p < 0.01$; ***, $p < 0.001$)

## 4. Discussion

In this study, we propose a framework that directly decodes semantic content of natural images from brain signals without leveraging visual information. Further analysis based on model not only provides further support for current neuroscience findings on visual semantic processing, but also reveals the neural mechanisms of brain's semantic representation engaged by distinct complex visual contents.

Our ROI-based analysis provides compelling evidence for the distributed nature of semantic representation and the critical roles of specific brain regions within the established semantic network(Huth et al., 2016). The MT+ complex is a key visual processing region associated with motion perception and object recognition(Kourtzi, Bülthoff, Erb, & Grodd, 2002; Maus, Fischer, & Whitney, 2013). The ventral stream is one of the two primary pathway in human visual system, which is essential for processing static object features and forming semantic concepts(Tyler et al., 2013). And the inferior parietal involves in cognitive functions such as attention, spatial cognition and semantic processing(Chou, Chen, Wu, & Booth, 2009; Seghier, 2013). These brain regions work synergistically to represent and integrate semantic information from complex visual scenes, thus serving as the primary contributors in semantic decoding. The SHAP analysis robustly reinforces these findings, highlighting the prominent contribution of these brain regions to model performance and emphasizing their central role in visual semantic representation. In addition, the supplementary validation on BOLD5000 dataset yielded a broadly similar pattern of ROI contributions, providing a more general principle of visual semantic processing.

Further analysis of model performance across semantic categories revealed nuanced aspects of neural representation and perceptual intensity. Our model effectively decoded the presence of semantic information related to person and motion, reflecting the functional specialization of brain regions for these semantic domains(Downing et al., 2001; Kanwisher et al., 1997). However its performance in reconstructing detailed and holistic semantic content did not significantly surpass other categories. This may reflect the inherent complexity of semantic representations associated with sports and motion, potentially encompassing richer spatio-temporal features and engaging broader neural networks beyond the visual cortex, even when presented in static images(Proverbio, Riva, & Zani, 2009). Meanwhile, the high decoding accuracy observed for categories like "zebra" and "airplane" may stem from their distinctive visual features, which are readily encoded and lead to unique neural patterns that facilitate the decoding process. Conversely, categories such as "backpack" and "suitcase" prove challenging to decode, as these objects are often appear in conjunction with human figures or contextually embedded within scenes, and their neural representations might be entangled with person- and building-related semantic features. Similarly, the poor decoding for small objects (e.g. "spoon", "cup") suggests potential attention selection biases in naturalistic visual processing. Analysis of the upper bound established with human-provided captions also indicate the inherent incompleteness in semantic retrieval for these categories, further reflecting the underlying mechanisms of attention selection and semantic processing of complex visual scenes without a specific task demand.

Differences in activation patterns between distinct semantic contents observed on different brain regions illuminate the neural instantiation of distributed and embodied semantic theories. The "living" categories showed reduced decoding performance compared to "non-living" categories. After accounting for decoding performance, caption length and lexical diversity, the "living" categories showed significant lower SHAP values across key ROIs compared to "non-living" categories, suggesting that recognition and representation of living things may rely on the more synergistic and distributed activation of broader neural networks(Zannino et al., 2010). In contrast, motion-related items demonstrated enhanced contributions from MT+ complex and dorsal stream visual regions. These areas, associated with motion perception and spatial processing, may represent objects with motion-related features or functions (e.g., graspability)(Almeida, Mahon, & Caramazza, 2010), in agreement with embodied semantic theories(Fernandino et al., 2016). Notably, we found that the ventral stream visual region, typically associated with static object recognition, also encode information about motion-related objects, implicating a more distributed network for encoding motion information. Moreover, the TPO junction, a region involved in higher-level cognition(De Benedictis et al., 2014), showed consistent contributions across semantic categories, reflecting its role in advanced functions like semantic integration, non-modal semantic representation and abstract conceptualization, rather than category-specific semantic processing.

To enhance semantic decoding of visual content from brain signals, we adopt a more straightforward method to directly generate textual descriptions without leveraging any visual information. Prior work utilizing image or its embedding as intermediary to reconstruct semantic information, therefore early visual cortices serving a key role as model input which encode basic perceptual features like shape, color and texture. However, in semantic reconstruction tasks, semantic information is predominantly represented and processed in broader brain regions and their functional connections within the semantic network. Therefore, in order to refine our understanding of the brain's semantic processing, future research should expand the scope of neuroimaging data beyond visual cortices. It's crucial to incorporate voxels from a broader network of brain regions that contribute to advanced cognitive functions, enabling illumination of how distributed semantic network collaboratively encode and integrate semantic content derived from complex visual scenes. Furthermore, theory-driven feature extraction informed by established neuroscientific principles enhances both efficiency and performance of deep learning models, as evidence by psychologically plausible models outperforming neural language models when mapping semantic representations to brain activation(Zhang, Wang, Dong, Yu, & Zong, 2024). The integration of psychological and neuroscientific insights promises to not only enhance the performance and interpretability of brain decoding models but also to uncover more ecologically valid neural representations of semantic information.

Compared to previous methods employing image captioning models or multimodal LLMs, our text-based decoding model yields lower scores on low-level metrics (e.g., BLEU-k, ROUGE, CIDEr), which assess the similarity in basic textual features (e.g., word choice and syntax). However, it achieved superior performance on high-level metrics (SPICE, CLIP-based scores) that evaluate semantic relevance and contextual fidelity. This dissociation underscores a critical point: while token-based text generation models excel at tasks emphasizing lexical overlap and syntactic accuracy, our sentence embedding-based method is better suited for capturing holistic semantic information and contextual coherence from brain signals. Despite minor deviations in syntax and wording compared to reference captions, our method recovers more accurate semantic information, particularly in tasks involving understanding and integration of complex visual and conceptual semantic contents. As a key component of our method, sentence embeddings encapsulate high-level conceptual information, serving as the bridge between raw neural signals and semantic interpretation. Future work should aggressively investigate alternative LLM architectures and embedding strategies, including exploring transformer-based models optimized for nuanced semantic capture and contextualized embeddings to enrich meaning representation. Critically, this line of inquiry offers the exciting potential to move beyond brain decoding models and facilitate the development of language models.

A critical methodological debate in neural decoding centers on the trade-off between decoding performance and interpretability(Kriegeskorte & Kievit, 2013). Traditional linear models are elegantly transparent, providing more detailed functional interpretations. In contrast, nonlinear deep learning models often achieve superior decoding performance by capturing complex spatial and hierarchical interactions, but normally fail to provide direct explicit mechanistic insights. To address this limitation, we utilized SHAP analysis to estimate the margin contribution of specific ROIs. However, we acknowledge certain limitations in the spatial granularity of our SHAP analysis. Our ROI parcellation was designed as a tractable compromise between anatomical granularity and computational feasibility. While an exact SHAP estimation on highly granular anatomical regions would yields more detailed functional interpretations, exact SHAP calculation scales exponentially with the number of features. Future studies could extend the present framework to a finer parcellation and bypass computational bottlenecks by using approximate SHAP extimators, such as Monte Carlo sampling methods or hierarchy-based SHAP approximation. These approaches would enable future decoding models to feasibly evaluate feature importance across highly granular, localized ROIs, further disentangling the overlapping contributions of complex visual cortex.

## 5. Conclusion

This study presents an significant advancement in brain decoding by demonstrating the feasibility of directly reconstructing visual semantic content in the form of textual descriptions from fMRI signals. Our findings validate and refine the current neuroscientific understanding of visual semantic processing, revealing nuanced neural representations for category-specific semantic. The proposed neurocomputational framework not only provides a methodology for probing the neural basis of cognition, but also offers insights for bridging neuroscience and artificial intelligence and advancing the development of language models.

## Acknowledgements

This work was supported by the Research Center for Brain Cognition and Human Development, Guangdong, China (No. 2024B0303390003); the Striving for the First-Class, Improving Weak Links and Highlighting Features (SIH) Key Discipline for Psychology in South China Normal University; Key-Area Research and Development Program of Guangdong Province (2019B030335001).

## Declaration of competing interest

The authors declare no competing interests.

## Code Availability

Our code is publicly available on GitHub (https://github.com/AllenFung/brain2text).

## Appendix

Appendix A. CLIP-S distribution across full test set.

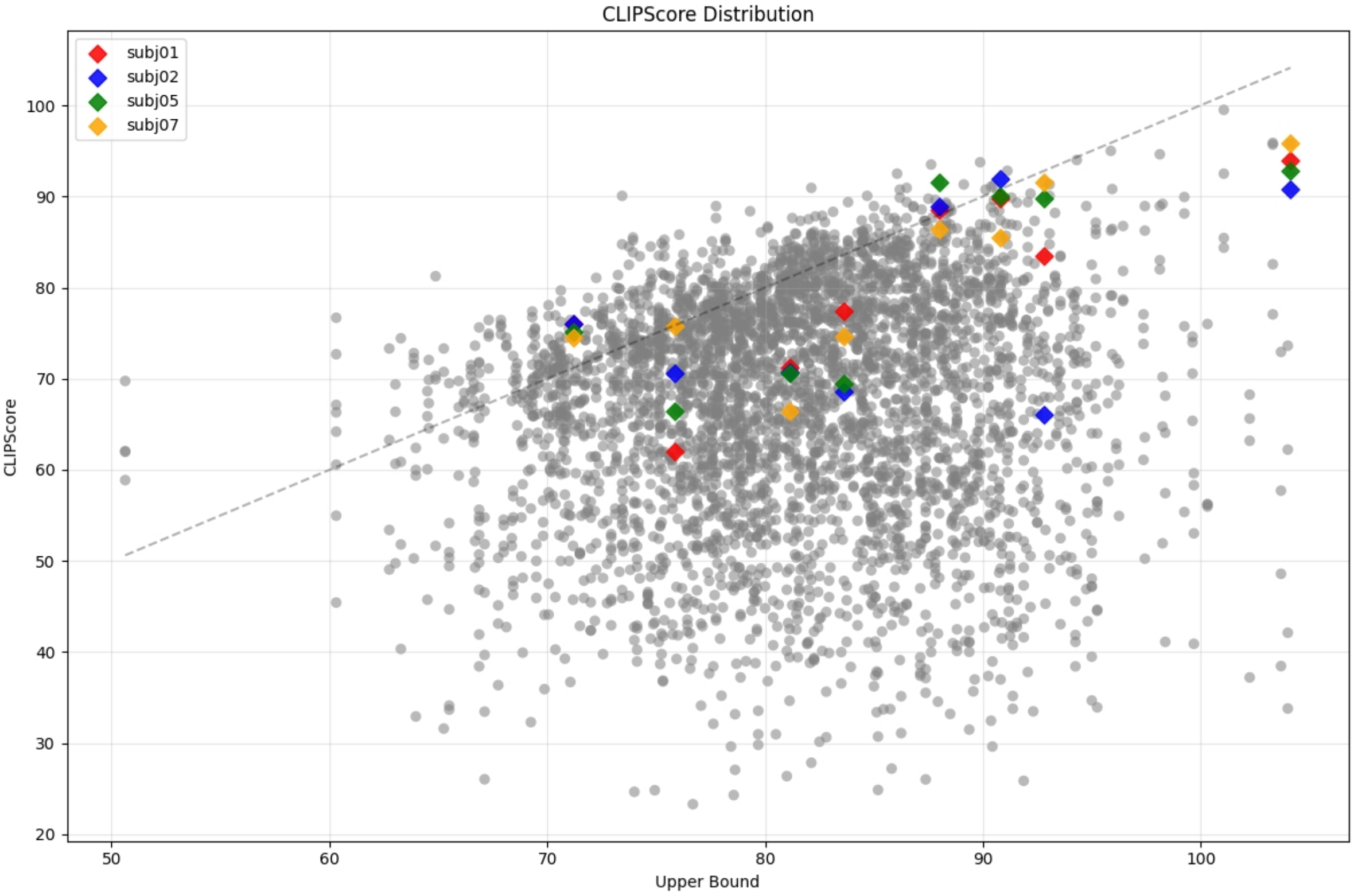

**Fig. A.1**. Distribution of decoding performance (CLIP-S) across across all four subjects (subj01, subj02, subj05, subj07) in the test set. The examples highlighted in Fig. 3 are annotated within this distribution to contextualize their representativeness relative to the overall decoding performance.

Appendix B. Contribution of different functional areas of visual cortex.

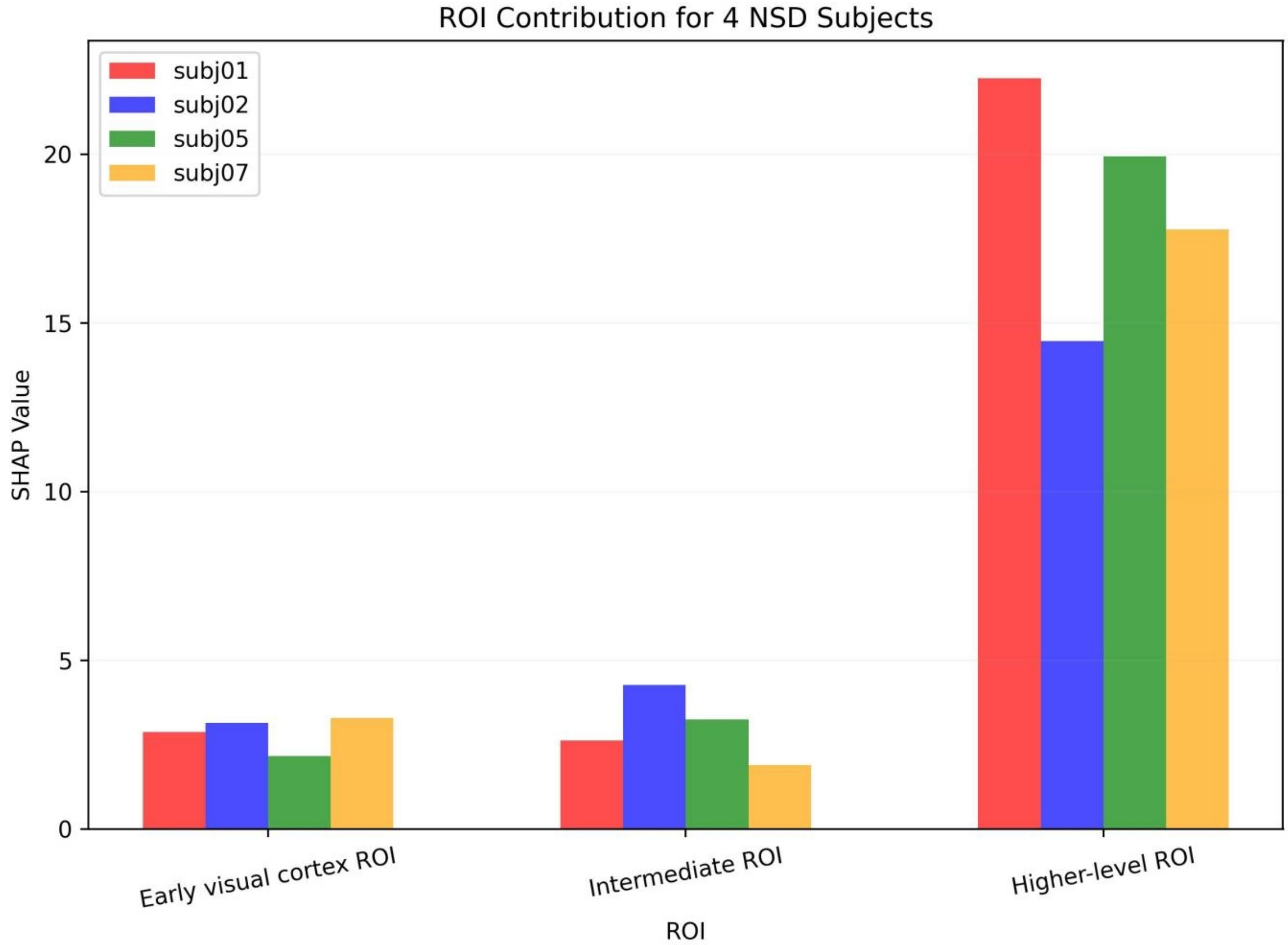

**Fig. B.1.** The contribution of the three broad ROIs from the visual cortex of all four subjects, which align with the model input of previous studies, to the semantic decoding performance.

## Appendix C

Table C.1 Evaluation results of different participants

| Subject ID | BLEU1 | BLUE2 | BLEU3 | BLEU4 | METEOR | ROUGE | CIDEr | SPICE | CLIP-S | RefCLIP-S | $R^2$(%) |
|---|---|---|---|---|---|---|---|---|---|---|---|
| NSD-subj01 | 56.79 | 36.47 | 21.85 | 13.07 | 21.56 | 43.20 | 47.84 | 14.01 | 69.14 | 74.07 | 34.17 |
| NSD-subj02 | 55.64 | 36.17 | 22.01 | 13.18 | 20.89 | 42.55 | 45.41 | 13.42 | 67.63 | 72.50 | 28.06 |
| NSD-subj05 | 55.07 | 35.63 | 21.70 | 13.25 | 20.53 | 41.83 | 45.39 | 12.96 | 66.94 | 71.63 | 27.65 |
| NSD-subj07 | 55.68 | 35.75 | 21.60 | 13.14 | 20.88 | 42.32 | 45.14 | 13.18 | 66.94 | 71.97 | 24.70 |
| BOLD5000-CSI1 | 35.77 | 17.30 | 6.52 | $4e^{-6}$ | 11.61 | 29.75 | 5.82 | 3.37 | 48.50 | 51.20 | 4.10 |
| BOLD5000-CSI2 | 36.52 | 15.47 | 4.90 | $3e^{-6}$ | 10.66 | 29.65 | 4.68 | 2.22 | 45.34 | 46.44 | 5.44 |
| BOLD5000-CSI3 | 37.03 | 18.42 | 8.38 | 4.59 | 12.11 | 31.04 | 9.01 | 4.07 | 47.00 | 51.60 | 4.64 |

Note that evaluation of BOLD5000 dataset was limited to the COCO subset, as only COCO images were provided with human-annotated ground truth captions. $R^2$ was computed as the mean voxel-wise GLM $R^2$ within higher-level ROIs across all scanning runs, reflecting the test-retest reliability of repeated trials and serves as an indicator of fMRI data quality.

## Appendix D. Cross-dataset validation on BOLD5000

To evaluate whether the results of SHAP analysis generalizes across datasets, we conducted an additional validation on the BOLD5000 dataset. The overall decoding and attribution pipeline followed the same procedure as in the main analysis, and the data split was kept consistent with prior studies to ensure comparability. For ROI definition, we used the "nsdgeneral" mask of fsaverage space provided in the NSD study, and mapped them to the participants' native volume space. This allowed us to extract higher-level visual cortex ROIs as model input that was consistent with the main procedure. Because BOLD5000 does not provide human-given captions paired with each image, we generated multiple synthetic captions using three different LLMs. Since we employed CLIP-S, which computes the semantic similarity between reconstructed text and ground truth images, for SHAP analysis as the evaluation metric, synthetic captions were only utilized during training phase and will not affect the results of SHAP analysis.

Appendix E. Results of hyperparameter tuning

| D \ L | 2 | 4 | 8 | 16 |
|---|---|---|---|---|
| w/o upe | 40.67 | 56.60 | 59.57 | 59.23 |
| 2 | 60.50 | 63.05 | 62.74 | 62.20 |
| 4 | 59.20 | 64.06* | 62.45 | 63.20 |
| 6 | 58.60 | 63.60 | 63.77 | 63.77 |

Results of model trained with different hyperparameter configurations, evaluated using CLIP scores. 'L', the number of latent channels in projected brain tokens; 'D', the number of layers in transformer architecture; 'upe', the universal perceive encoder (transformer).

Appendix F. HCP regions included in higher-level visual parcellation

| HCP region ID | HCP region | ROI parcellation | Voxel numbers (1.8mm native space) |
|---|---|---|---|
| 157 | FST | MT+ Complex | 320 |
| 20 | LO1 | MT+ Complex | 152 |
| 21 | LO2 | MT+ Complex | 247 |
| 159 | LO3 | MT+ Complex | 235 |
| 2 | MST | MT+ Complex | 228 |
| 23 | MT | MT+ Complex | 210 |
| 138 | PH | MT+ Complex | 624 |
| 158 | V3CD | MT+ Complex | 229 |
| 156 | V4t | MT+ Complex | 200 |
| 18 | FFC | Ventral Stream Visual | 678 |
| 22 | PIT | Ventral Stream Visual | 294 |
| 7 | V8 | Ventral Stream Visual | 282 |
| 153 | VMV1 | Ventral Stream Visual | 146 |
| 160 | VMV2 | Ventral Stream Visual | 207 |
| 154 | VMV3 | Ventral Stream Visual | 243 |
| 163 | VVC | Ventral Stream Visual | 391 |
| 146 | IP0 | Inferior Parietal | 292 |
| 145 | IP1 | Inferior Parietal | 74 |
| 150 | PGi | Inferior Parietal | 86 |
| 142 | PGp | Inferior Parietal | 491 |
| 151 | PGs | Inferior Parietal | 32 |
| 17 | IPS1 | Dorsal Stream Visual | 313 |
| 13 | V3A | Dorsal Stream Visual | 357 |
| 19 | V3B | Dorsal Stream Visual | 170 |
| 16 | V7 | Dorsal Stream Visual | 149 |
| 140 | TPOJ2 | TPO Junction | 221 |
| 141 | TPOJ3 | TPO Junction | 284 |
| 135 | TF | Medial Temporal | 45 |
| 126 | PHA1 | Medial Temporal | 106 |
| 155 | PHA2 | Medial Temporal | 106 |
| 127 | PHA3 | Medial Temporal | 165 |
| 5 | V3 | Secondary Visual | 1304 |
| 6 | V4 | Secondary Visual | 1147 |
| 137 | PHT | Lateral Temporal | 42 |
| 136 | TE2p | Lateral Temporal | 266 |
| 50 | MIP | Superior Parietal | 139 |